\documentclass[a4paper,11pt]{article}
\pdfoutput=1 

\usepackage{jinstpub} 

\usepackage[colorinlistoftodos]{todonotes}

\usepackage{epsfig}
\usepackage{amssymb}
\usepackage{lineno}
\usepackage{amsmath}
\usepackage{siunitx}
\usepackage{url}
\usepackage{booktabs}
\usepackage[utf8]{inputenc}
\usepackage{caption}  
\usepackage{subfig}

\newcommand{\cb}{CeBr$_3$(LB)}
\newcommand{\ly}{LYSO:Ce}

\usepackage{titlesec}
\titleformat{\paragraph}[runin]
{\bfseries\scshape}{\theparagraph}{1em}{}
\setcitestyle{square}

\date{\today}

\title{Characterisation of a \cb\ detector for space application}
\author[a,1]{A.~Di Giovanni,%
\note{Corresponding authors.}}
\author[a,1]{L.~Manenti,}
\author[a]{F.~AlKhouri,}
\author[a]{L.R.~AlKindi,}
\author[b]{A.~AlMannaei,}
\author[b]{A.~Al Qasim,}
\author[a]{M.L.~Benabderrahmane,}
\author[a]{G.~Bruno,}
\author[c]{V.~Conicella,}
\author[a]{O.~Fawwaz,}
\author[d]{P.~Marpu,}
\author[a]{P.~Panicker,}
\author[e,f]{C.~Pittori,}
\author[a]{M.S.~Roberts,}
\author[d]{T.~Vu,}
\author[a]{and F.~Arneodo}

\affiliation[a]{New York University Abu Dhabi, Saadyat Island, Abu Dhabi, UAE}
\affiliation[b]{University College London, London, UK}
\affiliation[c]{Università degli Studi Roma Tre, Rome, Italy}
\affiliation[d]{YahSat Space Lab, Khalifa University of Science and Technology, Abu Dhabi, UAE}
\affiliation[e]{ASI Space Science Data Center (ASI-SSDC), Rome, Italy}
\affiliation[f]{INAF, Osservatorio Astronomico di Roma (INAF-OAR), Monte Porzio Catone (Rome), Italy}

\emailAdd{laura.manenti@nyu.edu}
\emailAdd{adriano.digiovanni@nyu.edu}

\abstract{We describe the performance of a $\mathrm{23\times 23\times30 ~mm^3}$ low background cerium bromide, \cb, scintillator crystal coupled to a Hamamatsu R11265U-200 photomultiplier. This detector will be the building block for a gamma-ray detector array designed to be the payload for a CubeSat to be launched in 2020. The aim of the mission is to study flashes of gamma rays of terrestrial origin. The design of the detector has been tuned for the detection of gamma rays in the \SI{20}{\keV }--\SI{3}{\MeV} energy range.}


\begin{document}
\maketitle
\flushbottom

\section{Introduction}
As of June 2019, more than two-thousand CubeSats have already been launched into space~\cite{nanosatsWebsite}. CubeSats are miniature satellites consisting of one standard unit (1U) with dimensions of \linebreak
$\mathrm{10  \times 10 \times 10\,~ cm^3 }$, weighting about \SI{1.3}{\kilo\gram} and a power consumption of less than 2\,W. CubeSats range from as small as 1U to as big as 12U and may serve different purposes, such as scientific, educational, commercial, military etc. 
They were first proposed by Professor Jordi PuigSuari at California Polytechnic State University (Cal Poly) and Professor Bob Twiggs at Stanford University's Space Systems Development Laboratory (SSDL), whose original project, called CubeSat program, was to ``develop a picosatellite standard that significantly reduces the cost and development time of student satellites''~\cite{PuingSuari2001}. This work presents one of the four crystals of an array being planned to operate within 1U of a 3U CubeSat to be launched in 2020 to the International Space Station (ISS), from where it will be deployed into Low-Earth Orbit (LEO). 
The purpose of this CubeSat mission, temporarily called RAADSat (Rapid Acquisition Atmospheric Detector Satellite),  is to detect and study Terrestrial Gamma-Ray Flashes (TGFs).

TGFs are bursts of gamma rays of atmospheric origins (thunderstorms) and can usually be detected by low-Earth orbit satellites~\cite{Fishman1994, Smith2005, Agile2010, Briggs2013, Ursi2017}. They can also be sporadically observed by detectors on aircraft~\cite{Smith2011} and at ground level~\cite{Hare2016, Abbasi2018}.
TGFs were first discovered by the Burst and Transient Source Experiment (BATSE) on the Compton Gamma Ray Observatory (CGRO) in 1994~\cite{Fishman1994}. 
BATSE's primary objective was to study Gamma-Ray Bursts (GRBs), which are events  of much longer duration compared to TGFs. BATSE  only detected a small number of TGFs over its years of operation (about one event recorded every two months). In fact, the shortest time scale on which BATSE could trigger was \SI{64}{\milli\second}, almost three orders of magnitude larger than the average TGF duration, therefore only very bright TGFs could trigger on such a long time scale. The median TGF duration is less than \SI{100}{\micro\second}, and TGFs as short as \SI{20}{\micro\second} have been reported \cite{Agile_2015}.

Since their discovery \cite{Dwyer2012}, it was apparent that TGFs were correlated with thunderstorms, but it was only in 1996 that Inan and Reising~\cite{Inan1996} provided the first direct evidence that electrical activity was associated with TGFs. Intense radio burst signals of very low frequency (VLF), called ``radio atmospherics'' or ``sferics'' \cite{Connaughton_2013},   typical of natural lightning discharges, were measured at Palmer Station, Antartica, along the CGRO's footprint~\cite{Inan1996}.
In 2005, the Reuven High Energy Solar Spectroscopic Imager (RHESSI) satellite recorded 86 TGFs over six months~\cite{Smith2005}. 
The observed rate corresponded to $\approx$50 TGFs globally each day. Currently, assuming nearly dead-time free TGF detection, the expected TGF rate is of about 1500 events globally per day~\cite{Briggs_2013}. 

Even though TGF detection capabilities have substantially improved over the years, the exact mechanism behind TGF production is yet to be fully understood. 
The current leading model proposed to explain TGFs is the Relativistic Runaway Electron Avalanche (RREA).
For thunderstorms with electric fields above $2.84\times10^5\times n\,$V/m, where $n$ is the density of air with respect to  sea level, an electron gains more energy than what it loses in collisions with air, hence ``running away''. Runaway electrons were first theorised by Wilson in his seminal work of 1925~\cite{Wilson1925}, but it was only in 1992 that Gurevich, Milikh and Roussel-Dupr{\'e}~\cite{Gurevich1992} showed that elastic scattering between runaway electrons could initiate a multiplication process, also known as avalanche (apparently, Wilson was also aware of this process which he referred to as ``snow ball effect'' in his notes~\cite{Williams2010}). It should be noted that this is not what we conventionally call ``electric breakdown''~\cite{Dwyer2008b}, as the process is not self-sustained~\cite{Dwyer2011}. In fact, runaway electrons must be continuously provided by an external source 
(an exception to this is the ``cold runaway'', where thermal electrons normally created in thunderstorm discharges may all runaway if the electric field exceeds a critical value of \SI{30}{MV/m}~\cite{Dwyer2008b}).
The two other leading models of TGFs are instead self-sustained mechanisms~\cite{Dwyer2008,Skeltved2014}.One is called ``relativistic feedback'' model, and involves backward propagating runaway positrons and backscattered X-rays~\cite{Dwyer2012}; the other, instead, involves the acceleration of electrons in the strong electric field of stepping lightning leaders~\cite{Moss_2006, Carlson_2012, Celestin_2011}.

All missions before 2018 that have studied TGFs were actually not specifically designed for this scope. BATSE, AGILE (Astro‐rivelatore Gamma a Immagini LEggero), and the Fermi Gamma-Ray Space Telescope aimed at investigating cosmic gamma-ray sources, while RHESSI's primary goal was to explore solar flares by detecting photons from \SI{3}{\keV} to \SI{17}{\MeV}.
Past missions might have underestimated the intensity of these atmospheric phenomena because their time response was not optimised for these fast phenomena.

One major mission to study TGFs has already begun in 2018 with the installation of the ASIM (Atmosphere-Space Interactions Monitor) on the CEPF (Columbus External Platform Facility) of the ISS ~\cite{ASIM}. Another mission should start at the end of 2019 with the launch of TARANIS (Tool for the Analysis of Radiation from lightning and Sprites), a satellite designed to specifically observe TGFs and related atmospheric phenomena~\cite{webTARANIS}.
The RAADSat would be one of the first satellites built on a low-budget to study TGFs. 
The mission, whose cost is about half a million USD, is supported by the UAE Space Agency through the 2018 ``UAE Minisat Competition'', where the proposed payload has won the first place and has been awarded 100k USD.

\begin{figure}[t] 
\centering
\includegraphics[origin=c,trim= {1cm 0cm 1cm 0cm},width=0.6\textwidth, angle=0]{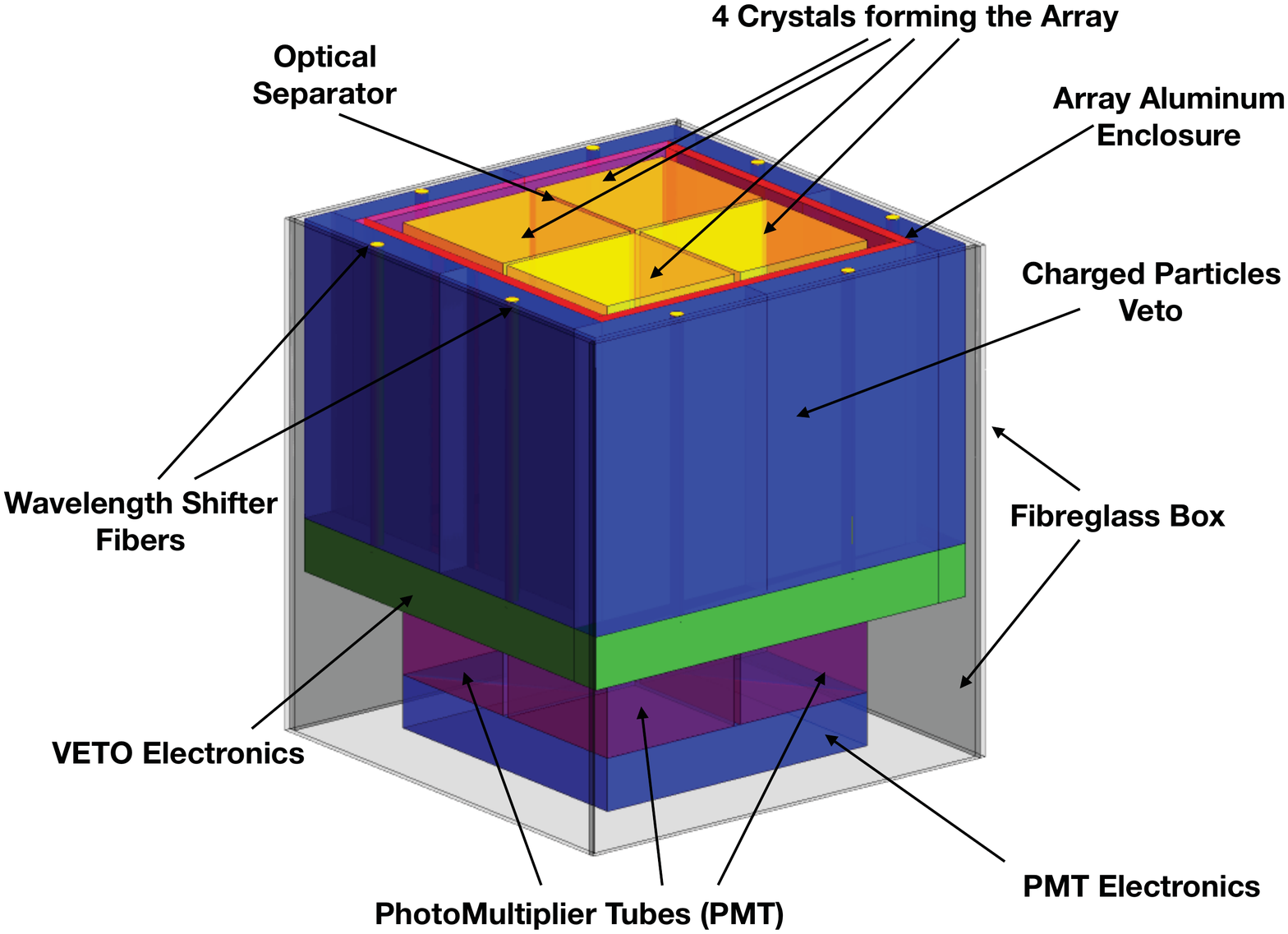}
   \caption{3D model of the $2\times2$ detector array used in the RAADSat. The four cerium bromide crystals (yellow) coupled to the respective photosensors (purple at the bottom) along with the electronic readout system (blue) are shown.}
   \label{fig:CubeSat}
\end{figure}

For this mission, we propose a fast detector which uses low background cerium bromide (\cb) scintillating crystals (manufactured by Scionix) coupled to fast, compact, and lightweight photomultiplier tubes (PMTs, model R11265U-200 manufactured by Hamamatsu).  The detector covers a wide energy range, from \SI{20}{\keV} up to \SI{3}{\MeV}.
Figure~\ref{fig:CubeSat} shows the layout of the detector, consisting of an array made of $\mathrm{2\times2}$ basic units, each separately operated and read out, together with an anti-coincidence detector to tag charged particle induced events. 

This work reports on the characterisation and preliminary tests of a basic space unit to assess its suitability to operate in space. 
The unit consists of one $\mathrm{23 \times 23 \times 30\, mm^3}$ \cb\ crystal coupled to one PMT, leading to an overall volume of $\mathrm{90  \times 90 \times 65 \,mm^3}$ and a total weight of $\approx$\SI{1}{\kg}.
Such configuration will fit into a single (1U) $\mathrm{10  \times 10 \times 10\, cm^3 }$ CubeSat unit. Our mission comprises of a 3U CubeSat bus, of which two units will house two almost identical detectors. The third will be dedicated to telemetry, transmission and necessary satellite control subsystems.

It is also worth mentioning that, albeit RAADSat was primarily designed for studying TGFs, it could also serve the HERMES (High Energy Rapid Modular Ensemble of Satellites) project. HERMES is a mission concept based on a swarm of picosatellites in low-Earth orbit to study hard X-ray/soft gamma-ray transients such as GRBs. One of the main objectives of the project is to test quantum space-time by measuring the time delay between GRB gammas of different energies~\cite{Luciano}. The picosatellites must be fast scintillators with high-accuracy time resolution. Both requirements are met by the RAADSat.

The paper is organised as follows: in section~\ref{sec:scintillator} the scintillator is described and the motivation behind its choice compared to other crystals. Section~\ref{sec:detector} describes the modification of the PMT voltage divider to comply with the power requirements of the future RAADSat mission. Section ~\ref{sec:characterisation} presents the measurement of the PMT gain at different high-voltage settings and the energy calibration and resolution characterisation by exposing the detector to a wide spectrum of radioactive gamma emitters. Lastly, section ~\ref{sec:characterizationVST} reports on the characterisation of this detector under vacuum at temperatures ranging from about \SI{-45}{\celsius} to \SI{50}{\celsius}.  

\section{The scintillator}
\label{sec:scintillator}
\begin{figure}[t]
    \centering
    \includegraphics[width=0.7\textwidth]{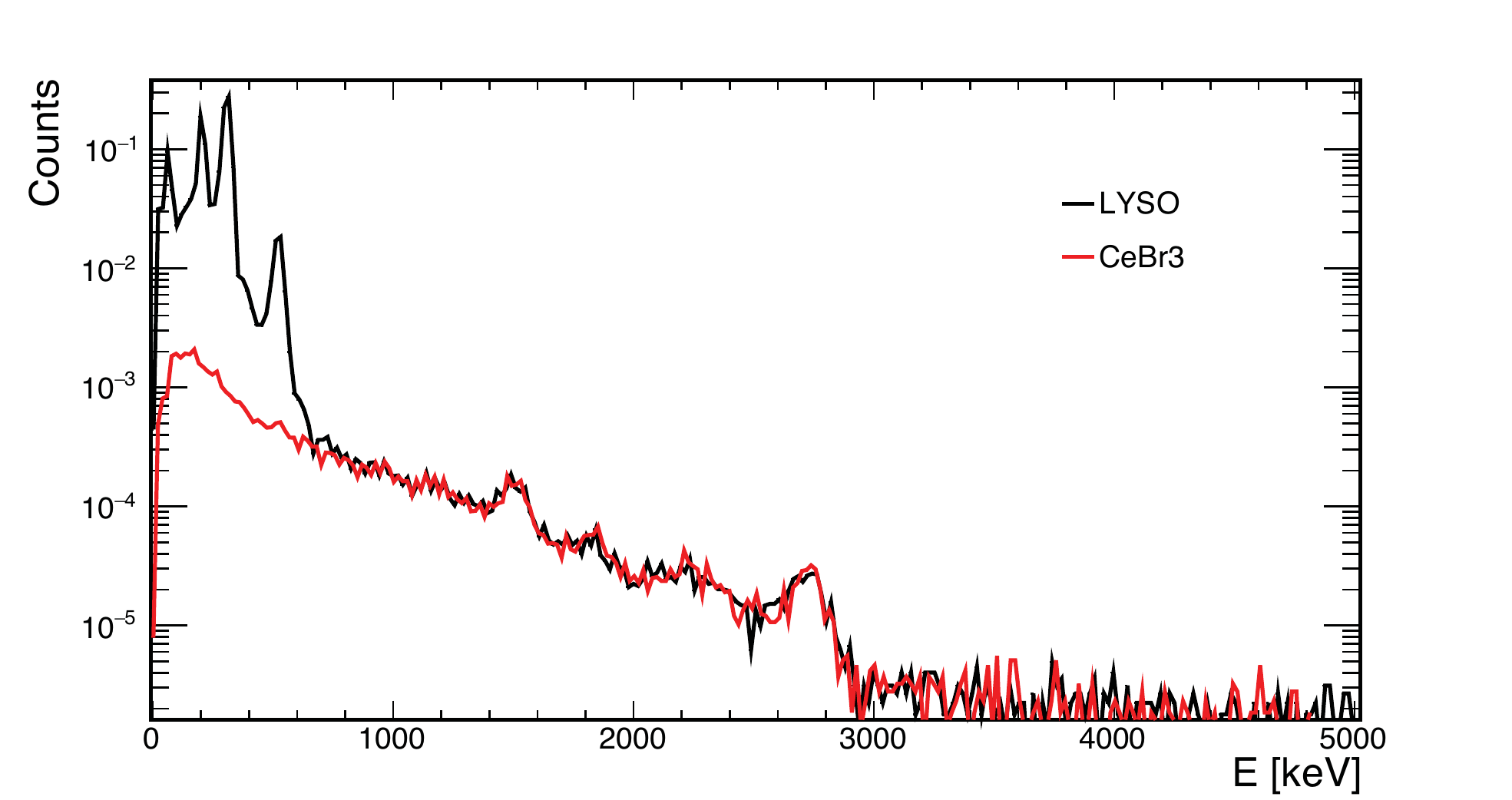}
    \caption{Activity spectrum for the \ly\ crystal and of the  \cb\ crystal used for this paper measured by a 3-inch ORTEC NaI:Tl detector. The estimated total activity of the \ly\ is about forty times higher than that of the \cb. All counts above 1000\,keV are ambient background rather than background due to the crystals and will not be present in space. The peak at $\approx$1.4\,MeV corresponds to the $^{40}$K and the one at $\approx$2.6\,MeV is the endpoint of the thorium decay chain.}
    \label{fig:intrBg}
\end{figure}

Given the timing characteristics of TGFs, a prerequisite for the detection medium is a short decay time to accommodate high event rates. 
A crystal which was initially considered as a viable option was the cerium-doped lutetium yttrium orthosilicate ($\rm Lu_{2(1-x)}Y_{2x}SiO_5$), also abbreviated as \ly. \ly\ has a relatively high density (\SI{7.25}{\gram\per\cubic\centi\metre}~\cite{Thiel2008}), good light yield ($\approx$5600--7600\,ph/MeV~\cite{Thiel2008}), good energy resolution, and a decay time of about \SI{40}{\ns} at \SI{420}{\nm}~\cite{Thiel2008}, with the advantage of not being hygroscopic~\cite{Chen2007}.  However, \ly\ is radioactive because of the presence of $^{176}$Lu, which decays via $\upbeta^-$ emission. Measurements performed with a NaI:Tl detector (shown in Fig.\ref{fig:intrBg}) resulted in an estimated total activity forty times higher than that of the \cb. For this reason, \ly\ was then discarded.

 \begin{figure}[ht]  
     \centering
     \includegraphics[width=0.3\textwidth]{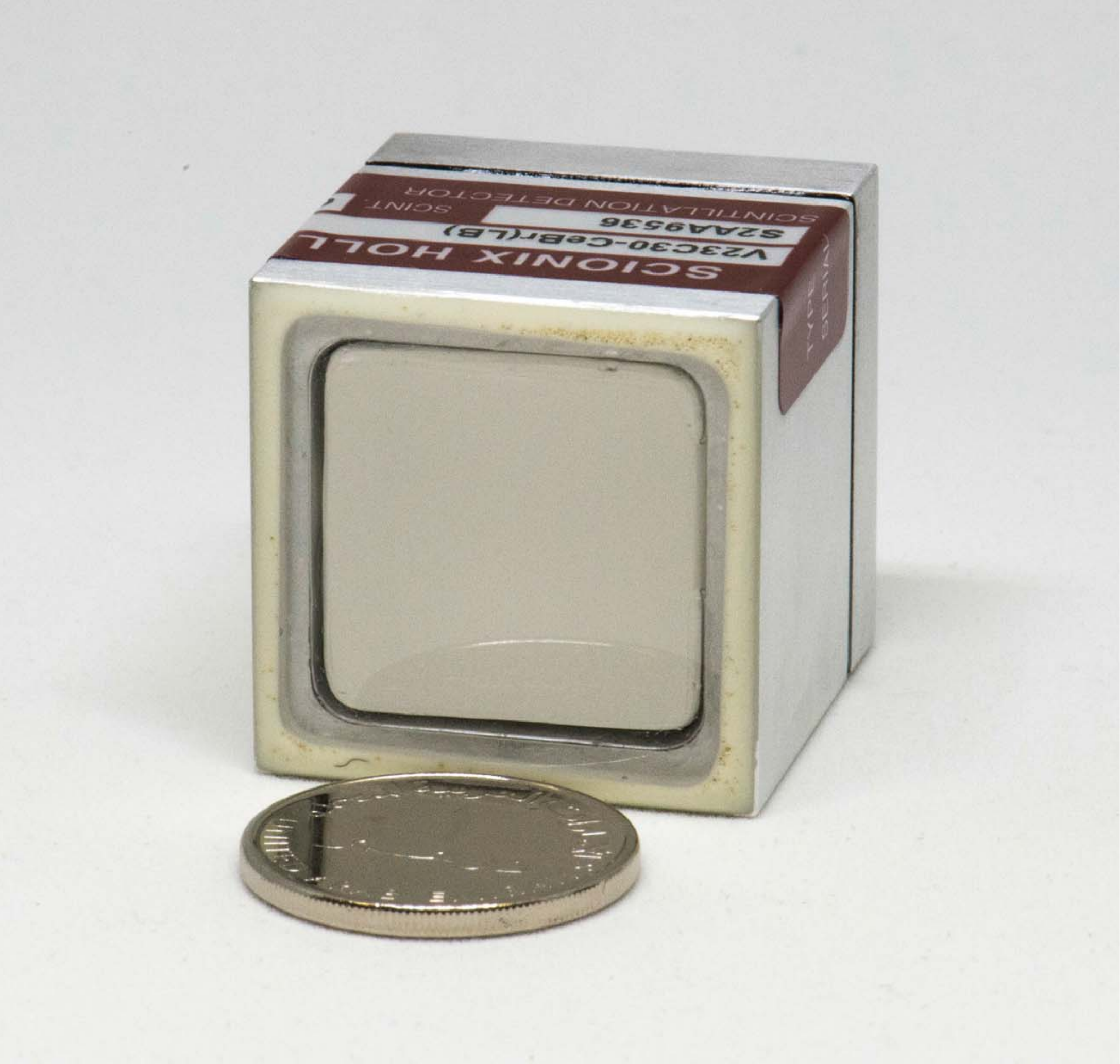} 
     \caption{The $\mathrm{23\times23\times30\,mm^3}$ \cb\ crystal by Scionix used in the present work shown next to one UAE Dirham coin.}
     \label{fig:crystal}
 \end{figure}

Another crystal that in recent years has attracted a lot of attention for space-based applications is \cb. This crystal has a relatively short decay time ($\approx$\SI{20}{\ns}) and, contrary to \ly, \cb\ shows a much lower intrinsic radioactivity~\cite{Quarati2013}. 

\begin{table}[b]
\centering
\caption[]{The table summarises the main properties of a traditional NaI:Tl, a LYSO:Ce, and \cb\ crystals. The values regarding the NaI:Tl are taken from~\cite{Knoll,Saint-Gobain}; the ones for \ly\ are from \cite{Thiel2008} and \cite{LYSOactivity}; the numbers for \cb\ are as reported in \cite{Quarati2013}.}
    \smallskip
    \begin{tabular} {@{}llll@{}}
    \toprule
     & NaI:Tl & \ly &  \cb\\   
    Density [g/cm$^3$]              & 3.67 & 7.25          & 5.18 \\
    Absolute light yield [photons/MeV]       & $\approx$38000 & $\approx$28000                  & $\approx$60000 \\
    Decay time [ns]                 & $\approx$230 & $\approx$40            & $\approx$20 \\
    Emission Peak [nm]              & 415 & 420           & 370 \\
    Hygroscopic                     & yes & no            & yes\\
    \end{tabular}
    \label{tab:comp} 
\end{table}
Although \cb\ is  less dense than \ly\ (see Table\ref{tab:comp} for comparison), resulting in a reduced stopping power, its crystal structure represents a promising material with a very low intrinsic radioactivity and an excellent photon yield ($\approx$60000\,photons/MeV~\cite{Quarati2013}). Hence, we decided to opt for the \cb, a sample of which we sourced from Scionix, Ltd (Netherlands). The sample, shown in Fig.~\ref{fig:crystal}, has dimensions of  $\mathrm{23\times23\times30\,mm^3}$ and is encapsulated in an aluminium case with overall dimension of $\mathrm{32\times32\times34\,mm^3}$. The crystal comes with an optical interface---a \SI{2}{\mm} thick quartz window---for protection. 

According to the manufacturer, the \cb\ emission spectrum peaks at \SI{380}{\nm} with a typical decay time of 18--\SI{20}{\ns}. The peak emission reported by~\cite{Quarati2013} is instead at \SI{370}{\nm}, while the decay time reported in the same work varies with the size and thickness of the crystal. 
The scintillation decay time of the crystal used in this work is 23.53$\pm$\SI{0.01}{\ns} and it has been measured by fitting a sample of 10000 waveforms generated by exposing the detector to a $\mathrm {^{54}Mn}$ source and acquired by using a LeCroy HDO6104 oscilloscope.

\section{The detector unit}
\label{sec:detector} 

\begin{figure}[h!] 
   \centering
   \includegraphics[width=0.3\textwidth]{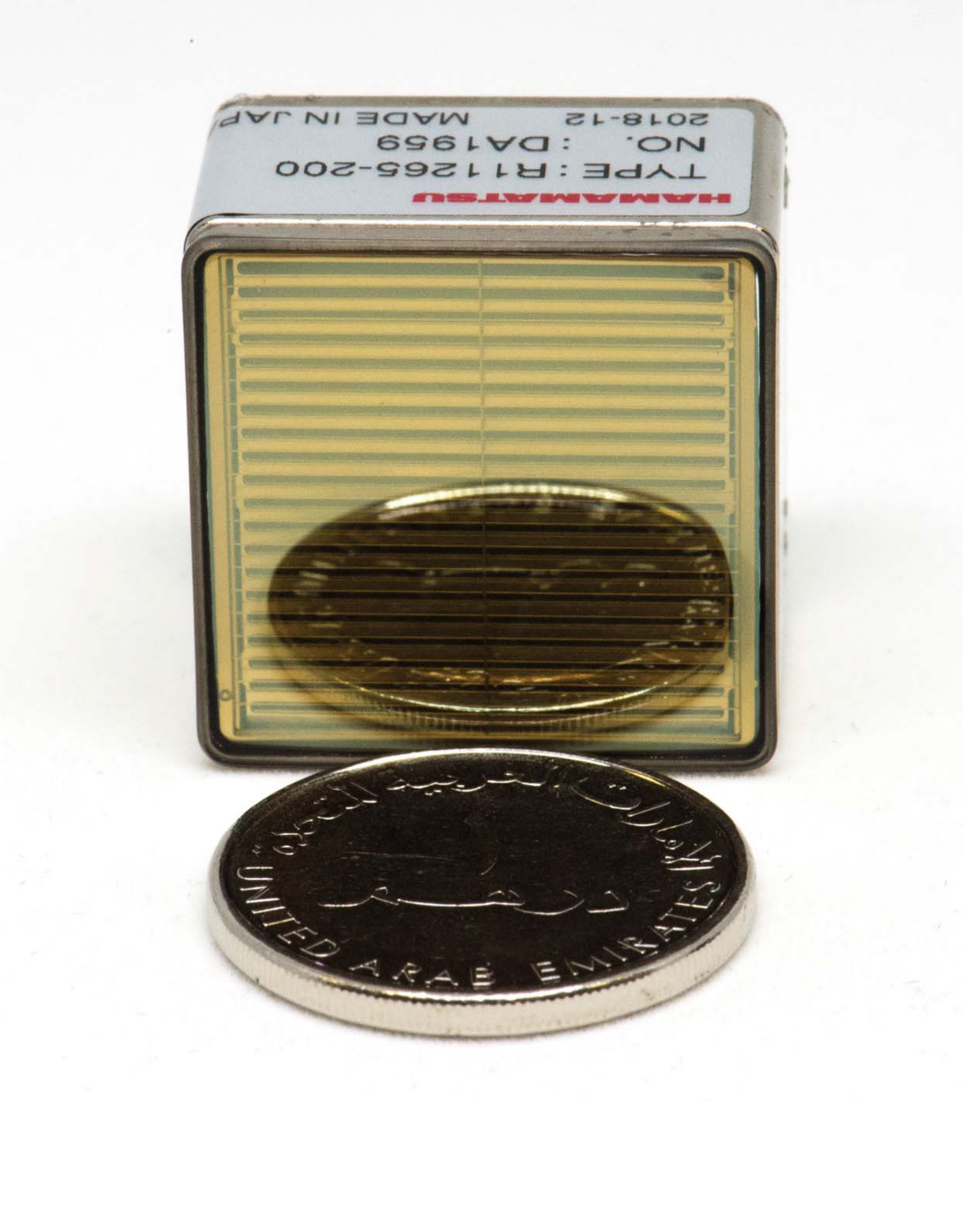} 
   \caption{The Hamamatsu photomultiplier R11265U-200 used in the present work, next to one UAE Dirham coin. }
   \label{fig:pmt}
\end{figure}

We looked for a compact, rugged photomultiplier tube to be coupled to the \cb\ crystal. The choice fell on the Hamamatsu R11265U-200 series, a sample of which is shown in Fig.~\ref{fig:pmt}.
 The area of the photocathode window, $23\times23$\,mm$^2$, matches the quartz window of our \cb\ crystal, and its limited weight of \SI{31}{\g} makes this tube a promising choice for our CubeSat application. The main characteristics of the PMT R11265U-200 are listed in Table~\ref{tab:pmt}.

\begin{figure}[h!] 
   \centering
   \includegraphics[width=0.4\textwidth, clip=true]{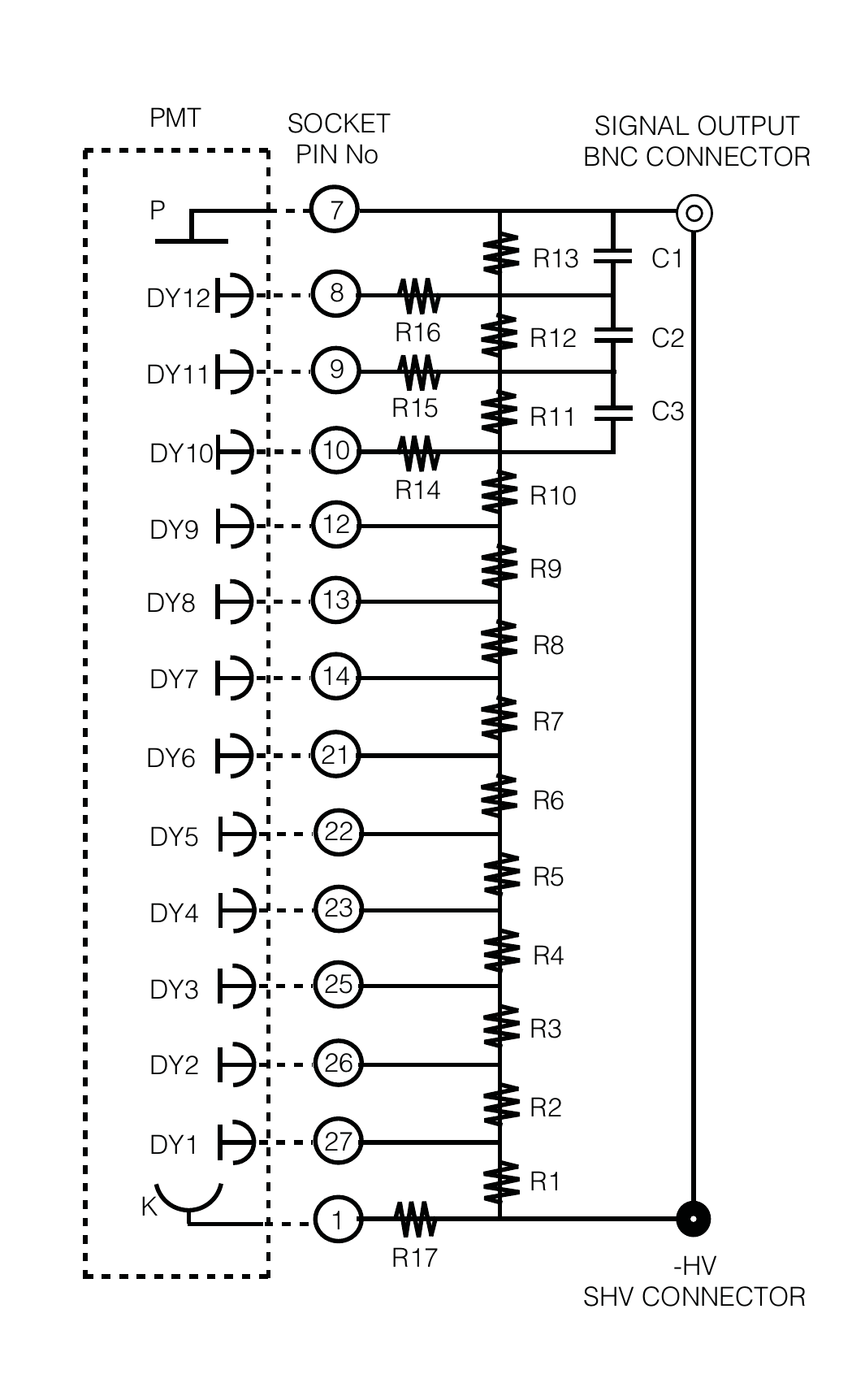}
   \caption{The voltage divider of the Hamamatsu R11265U-200 series. See Table \ref{tab:divider} for the values of all components. }
   \label{fig:divider}
\end{figure}

\begin{table}[b]
\centering
\caption[]{The table summarises the main characteristics of the PMT R11265U-200 chosen for our setup.}
    \smallskip
    \begin{tabular} {@{}lr@{}}
    \toprule
            Dimensions [mm$^3$]     & $23 \times 23 \times  30$ \\
            Photocathode              & Ultra-bialkali\\
            Weight [g]              & 31\\
            Peak sensitivity [nm]   & 400 \\
            Quantum efficiency &    $\approx$ 43 \% \\
    \bottomrule
        \end{tabular}
        \label{tab:pmt}
    \end{table}
The R11265U-200 PMT provided by Hamamatsu comes with a standard voltage divider (shown in Fig.~\ref{fig:divider}) with a total resistance of \SI{2.78}{\mega\ohm}. At an operating voltage of \SI{600}{\volt}, the total power consumption is \SI{130}{\milli\watt}. For a $2\times2$ array, the total required power (due to the voltage dividers only) would be incompatible with the overall power budget of a CubeSat. We have therefore assembled an alternative voltage divider with a total resistance of \SI{27.8}{\mega\ohm}, which hereafter will be referred as the ``10X divider''. 
The power consumption of the new divider is only \SI{13}{\milli\watt}.
However, the current flowing through the divider should be kept large compared to the current flowing from cathode to anode through the dynodes chain. Especially  in correspondence of intense flashes and high rates, the internal current at the peak of a pulse might become comparable to the divider current, causing the voltage between two successive dynodes to deviate from the nominal value and the PMT gain to drift. This effect is particularly important for the dynodes closer to the anode---where the current flowing from one dynode to the other is higher. To avoid PMT gain instability and a nonlinear response, it is common practice to add stabilising capacitors to the last stages of the resistor chain. Such capacitors supply the PMT with an electric charge when the signal pulse is formed, keeping the voltage drop between the dynodes constant. 
The values of the capacitors placed in parallel in the new divider layout have been modified according to the empirical relation~\cite{HamamatsuHandbook}
\begin{equation}
C \simeq \frac{I\times \Delta t}{V},
\end{equation}
where \textit{C} is the value of the stabilising capacitor, \textit{I} is the current associated to the signal pulse, $\Delta t$ is the pulse width, and \textit{V} the voltage across the capacitor. Considering a typical \SI{661}{\keV} gamma absorption in the crystal, with an average current pulse of the order of \SI{1}{\mA}, a width of \SI{40}{\nano\second} and \SI{40}{\V} across the capacitor, a capacitance value of $\approx$\SI{1}{\nano\farad} is obtained. Following a conservative approach, a capacitance of \SI{10}{\nano\farad} has been used for C1, C2, and C3 (see Fig.~\ref{fig:divider}).
The performance of the R11265U-200 PMT with the factory voltage divider and the 10X divider has been evaluated and the results are reported in section \ref{sec:characterisation}. 

\begin{table}[th]
    \centering
    \caption{Component values for the factory voltage divider and the 10X one, used in all the measurements of this work. See Fig. \ref{fig:divider} for the divider schematics. }
    \smallskip
    \begin{tabular}{ l | r r }
         \hline
         &Factory Divider & 10X Divider \\   
        \hline
        R1&        \SI{500}{\kilo\ohm} &\SI{5}{\mega\ohm} \\
        R2&        \SI{260}{\kilo\ohm} &\SI{2.6}{\mega\ohm} \\
        R5, R6&    \SI{160}{\kilo\ohm} &\SI{1.6}{\mega\ohm}\\
        R7 to R12& \SI{200}{\kilo\ohm} &\SI{2}{\mega\ohm}\\
        R13&       \SI{100}{\kilo\ohm} &\SI{1}{\mega\ohm}\\
        R14 to R16&\SI{51}{\ohm} &\SI{51}{\ohm}\\
        R17&       \SI{1}{\mega\ohm} &\SI{1}{\mega\ohm}\\
        C1 & \SI{10}{\nano\farad} & \SI{10}{\nano\farad} \\
        C2 & \SI{10}{\nano\farad} & \SI{10}{\nano\farad} \\
        C3 & \SI{10}{\nano\farad} & \SI{10}{\nano\farad} \\
        R$_{\rm TOT}$ &  \SI{2.78}{\mega\ohm} &\SI{27.8}{\mega\ohm}\\
        Current at V= -1000 V & \SI{359.7}{\micro\ampere} & \SI{36}{\micro\ampere} \\
        \hline
    \end{tabular}
    \label{tab:divider}
\end{table}

\section{Detector characterisation at room temperature}
\label{sec:characterisation}

 \begin{figure}[h!] 
   \centering
   \includegraphics[width=0.3\textwidth, clip=true]{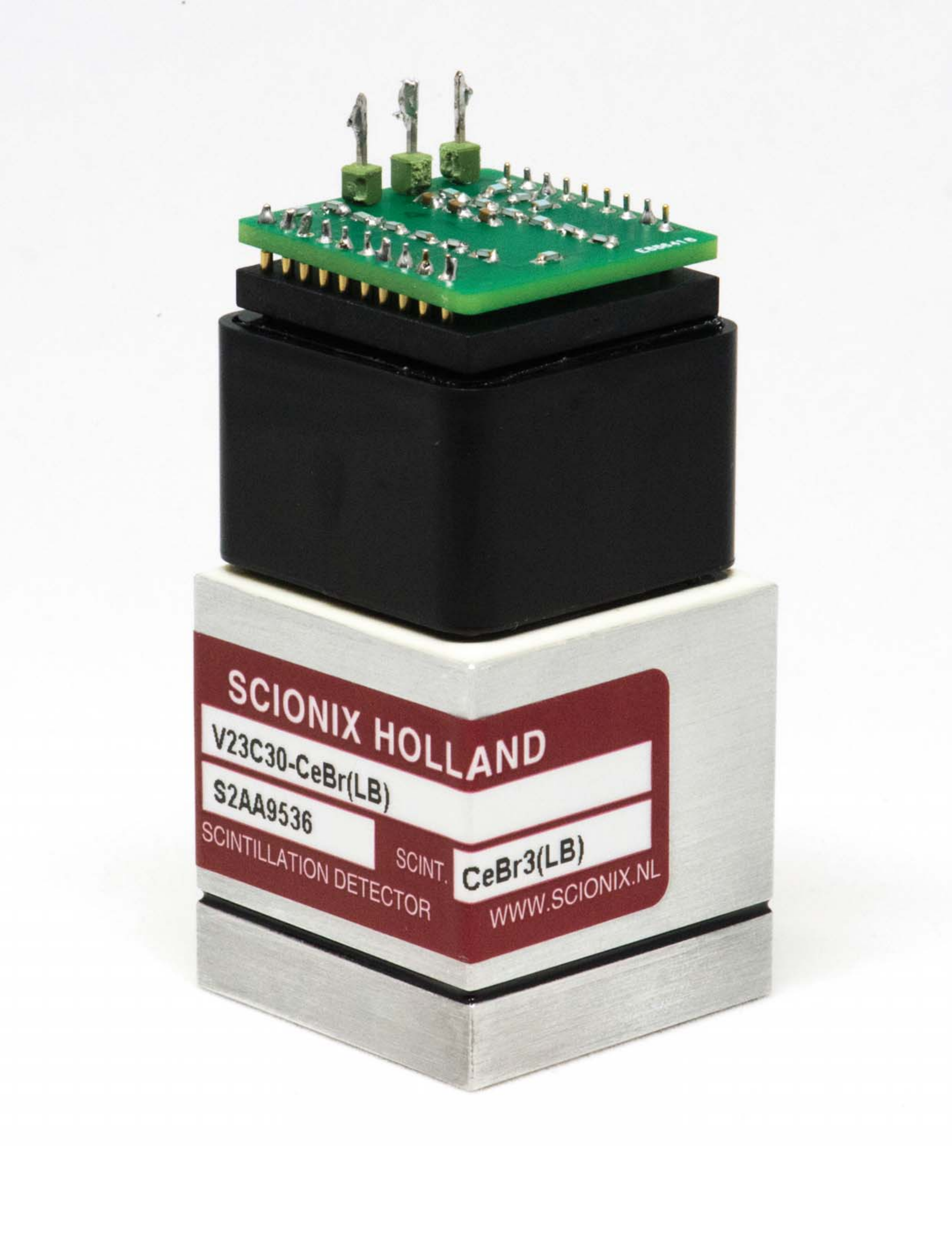}
   \caption{The assembly of the \cb\ crystal and  R11265U-200
   with the 10X divider.}
   \label{fig:detect}
 \end{figure}
The basic detector unit with the crystal coupled to the PMT is shown in Fig.~\ref{fig:detect}. In the final configuration, the unit will be inserted in a high strength silicone rubber case (Momentive RTV615) for two main reasons: firstly, to prevent high voltage discharges that could still be triggered by residual outgassing from the detector components even when in orbit. Secondly, the rubber acts as a shock absorber damping the vibrations during the launch phase. 

The gain of the R11265U-200 Hamamatsu photomultiplier equipped with the 10X divider has been measured as a function of the high voltage (HV). The PMT is lit by a pulsing LED and for each waveform acquired the pulse area is measured over a predefined region (such region is chosen so that the start and the end of all pulses fall within the chosen time window). The intensity of the LED is such that only zero, one or two photoelectrons are emitted from the photocathode at a time. To correct for baseline fluctuations, the integral over a ``pre-trigger'' region is also calculated and subtracted to the aforementioned area. Such measurements have all been performed using the LeCroy HDO6104 oscilloscope, whereas all subsequent fits have been done offline using the ROOT framework. 
Dividing the signal area by the coupling resistance of the oscilloscope (\SI{50}{\ohm}) effectively gives the charge collected by the PMT. 

\begin{figure}[t] 
    \centering
    \includegraphics[width= 0.8 \textwidth]{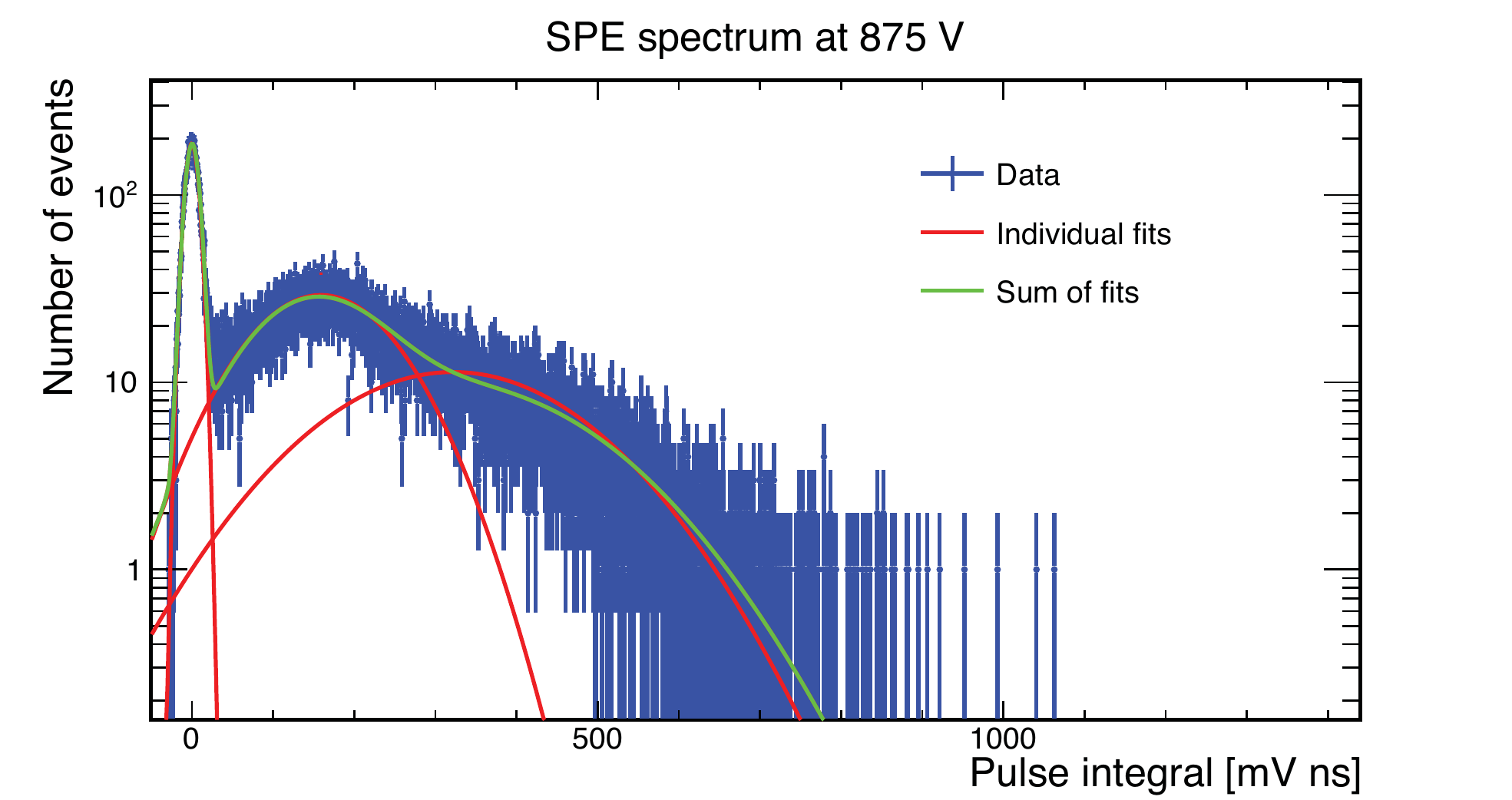}
    \caption{The distribution of the integral of the pulses in mV$\cdot$ns. The first peak is the pedestal, the second peak corresponds to single photoelectron events, the third to double photoelectrons events. The individual Gaussian fits are shown in red. The cumulative fit is shown in green.} 
    \label{fig:SPE}
\end{figure}

\begin{figure}[h!] 
    \centering
    \includegraphics[width= 0.8 \textwidth ]{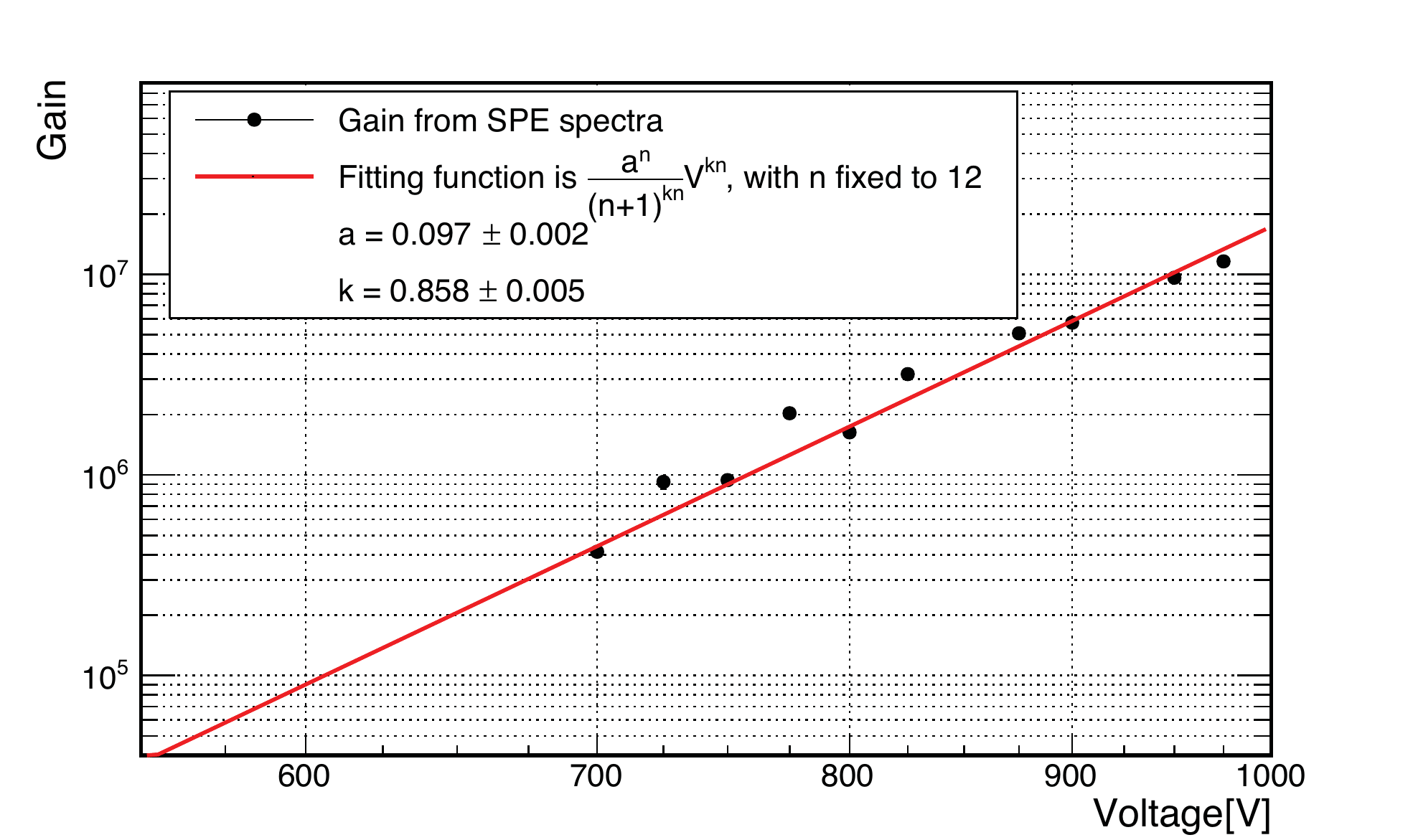}
    \caption{Gain measurement of the R11265U-200 photomultiplier tube equipped with the ``10X voltage divider'' as a function of the applied voltage bias. At \SI{600}{\V}, the gain, as calculated from the fitting function evaluated at \SI{600}{\V}, is $(9.0\pm2.8)10^{4}$.}
    \label{fig:GainPMT}
\end{figure}

The distribution of the integral of the pulses in mV$\cdot$ns is shown in Fig.\ref{fig:SPE}. The first peak is the pedestal and is associated to background processes (e.g. leakage current, thermal emission and electronic noise). The second peak comes from single photoelectrons and the third from double photoelectrons. The characteristic Gaussian shape is due to the multiplication process of the dynode chain. The difference between the mean of the single photoelectric peak $\mu_1$ and the pedestal $\mu_0$ in pC\footnote{$\mu_1$ and $\mu_0$ as obtained from the fit are in mV$\cdot$ns. To convert to pC, they must be divided by \SI{50}{\ohm} }, divided by the electron charge, $Q_{\rm e}$, gives the gain $G$ of the PMT,
\begin{equation}
    G = \frac{\mu_1-\mu_0}{Q_{\rm e}}.
\end{equation}    
By changing the HV applied to the PMT, the gain as a function of the high-voltage in Fig.~\ref{fig:GainPMT} is obtained. The points have been fitted with the following formula~\cite{HamamatsuHandbook}
\begin{equation}
    G = A{V^{kn}},
\end{equation}
where $n$ is fixed and corresponds to the number of dynodes (12 for the R11265U PMT series) and $A$ can be written as 
\begin{equation}
    A = \frac{a^n}{(n+1)^{kn}}.
\end{equation}
In Fig.~\ref{fig:GainPMT} $n$ was fixed to 12, with $k$ and $a$ being the only two free parameters of the fit. 

\begin{figure}[t] 
    \centering
    \includegraphics[angle=90, width=0.8\textwidth]{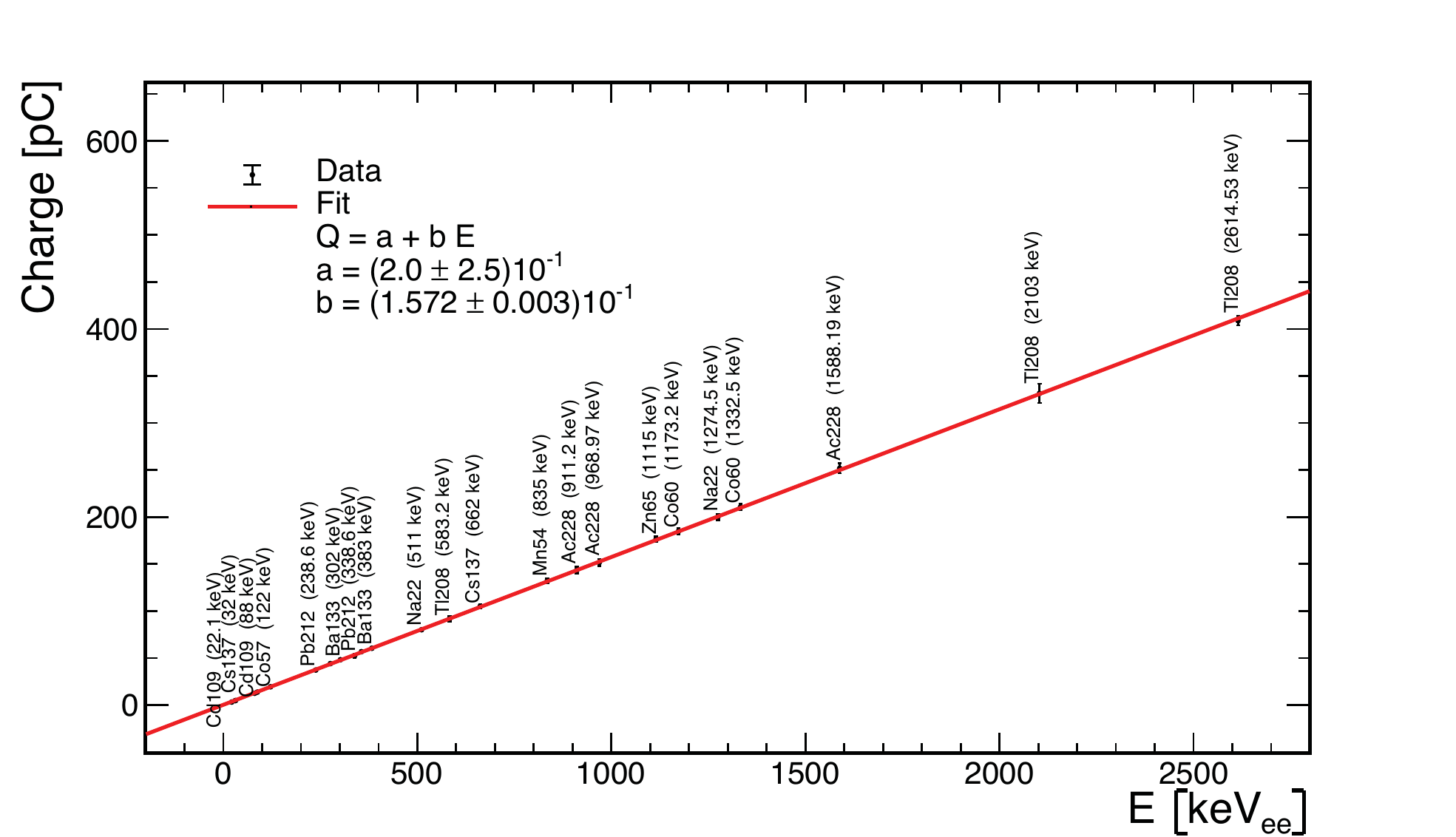}
    \caption{Calibration of the detector with a standard set of $\gamma$-ray sources and the PMT operated at 600\,V. The error bars correspond to the standard deviations of the individual energy peaks, that have been fitted with a Gaussian distribution. For each point the corresponding element and nominal energy in keV are shown. For clarity, the labels of the 31\,keV, 81\,keV, and 356\,keV data points from $^{133}$Ba have been removed.}
    \label{fig:calibration}
\end{figure}

\begin{figure}[h!]
    \centering  
    \includegraphics[angle=90,width=0.8\textwidth ]{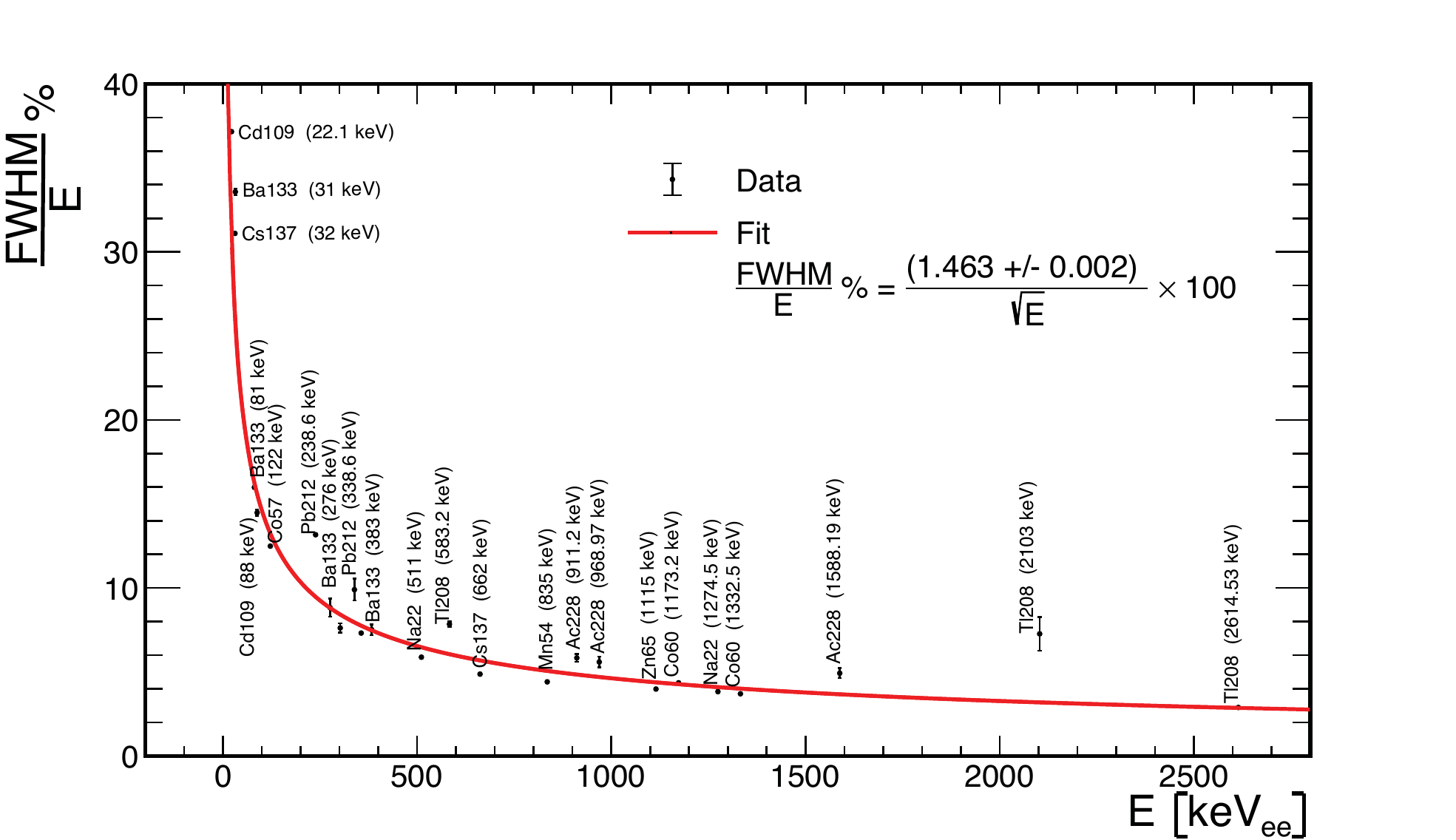}
    \caption{Energy resolution of the detector as a function of energy. The error bars are calculated by propagating the errors on the mean and the standard deviation of the individually fitted energy peaks.  For clarity, the labels of the 302\,keV and 356\,keV data points from $^{133}$Ba have been removed. Note that the resolution is highly dependent on the PMT operating bias, here at 600\,V.}
    \label{fig:resolution}
\end{figure}

We performed an energy calibration with a standard set of radioactive isotopes ($^{109}$Cd, $^{133}$Ba, $^{137}$Cs, $^{60}$Co, $^{57}$Co, $^{22}$Na, $^{54}$Mn, $^{65}$Zn, $^{232}$Th) ranging from \SI{22}{\keV} (from $^{109}$Cd) to \SI{2.6}{\MeV} (from $^{208}$Tl, the last decay of $^{232}$Th decay chain). Data have been taken at different high voltage values, namely \SI{600}{\V}, \SI{700}{\V}, \SI{800}{\V}, and \SI{900}{\V}. Here we only report the calibration and the corresponding energy resolution at \SI{600}{\V}. This will be the operating voltage during the space mission, as it is a good trade-off between performance and low power requirements. 
The calibration is shown in Fig.~\ref{fig:calibration} and the resolution, which varies from $\approx$30\% at $\approx$30\,keV  to less than 2\% at \SI{2.6}{\MeV}, is shown in Fig.~\ref{fig:resolution}.

Note that as the calibration uses gamma isotopes, the energy is in units of keV$_{\rm ee}$ (``electron equivalent''). For the same amount of energy deposited in the crystal, neutrons and alpha-particles yield less scintillation light, some being lost in non-radiative collisions. Hereby, we express the actual energy deposited by neutrons and alpha-particles in units of keV$_{\rm ne}$ (``nuclear equivalent''), with $\rm keV_{ne}>keV_{ee}$. We plan to measure the linear relation between keV$_{\rm ee}$ and keV$_{\rm ne}$ in the near future using a neutron source. 

\section{Detector characterisation as a function of temperature}
\label{sec:characterizationVST}

When orbiting Earth, the RAADSat will be exposed to range of temperatures from about \SI{-20}{\celsius} to \SI{40}{\celsius}. For this reason, we tested the basic detector unit between \SI{-45}{\celsius} and \SI{50}{\celsius} under vacuum. All the measurements have been performed in the thermal vacuum chamber at the YahSat Space Lab at Masdar Institute (Khalifa University). This chamber works at high vacuum ($\approx$$\mathrm{10^{-6}}$\,mbar) and allows the radiative thermal environment to be controlled within \SI{1}{\celsius}.
Three thermosensors (Pt100) were used: one attached to the PMT, one to the crystal and one to the platform where the PMT was seated on  the chamber. Temperature readings from the sensors were constantly monitored by a Raspberry PI. Before starting a measurement, we made sure to wait enough time (around one hour) for the temperature to stabilise. The temperatures reported in the following plots refer to the one measured by the thermosensor on the crystal: the temperature of the PMT was found compatible with such temperature within \SI{1}{\celsius}.

\begin{figure}[t]
    \centering  
    \includegraphics[width=0.7\textwidth ]{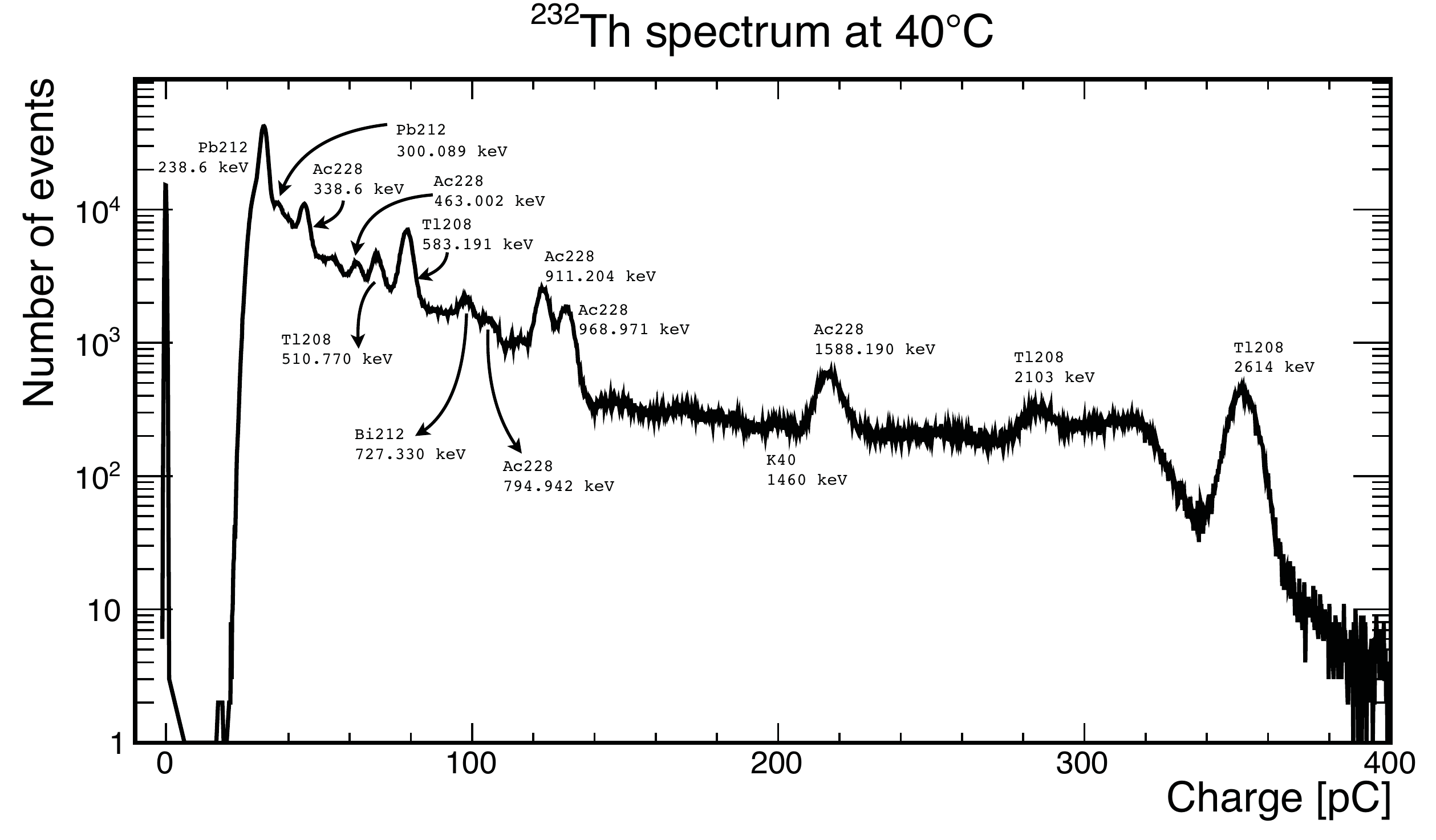}
    \caption{$^{232}$Th spectrum at \SI{40}{\celsius}, as measured in the thermal vacuum chamber at the Masdar Institute. The peaks have been labelled according to~\cite{international2013iaea}. Note that only the main peaks have been used in the energy calibration plot in Fig.~\ref{fig:Masdar_QvsE}.}
    \label{fig:232Th_spectrum}
\end{figure}

Firstly, we performed an energy calibration as a function of temperature using $^{232}$Th. The radioactive source was sealed in between two blank flanges and placed in front of the crystal side of the detector in the vacuum chamber. As an example, we show the spectrum taken at \SI{40}{\celsius} in Fig.~\ref{fig:232Th_spectrum}. The peaks have been labelled according to~\cite{international2013iaea}.

\begin{figure}[t]
    \centering  
    \includegraphics[width=0.8\textwidth ]{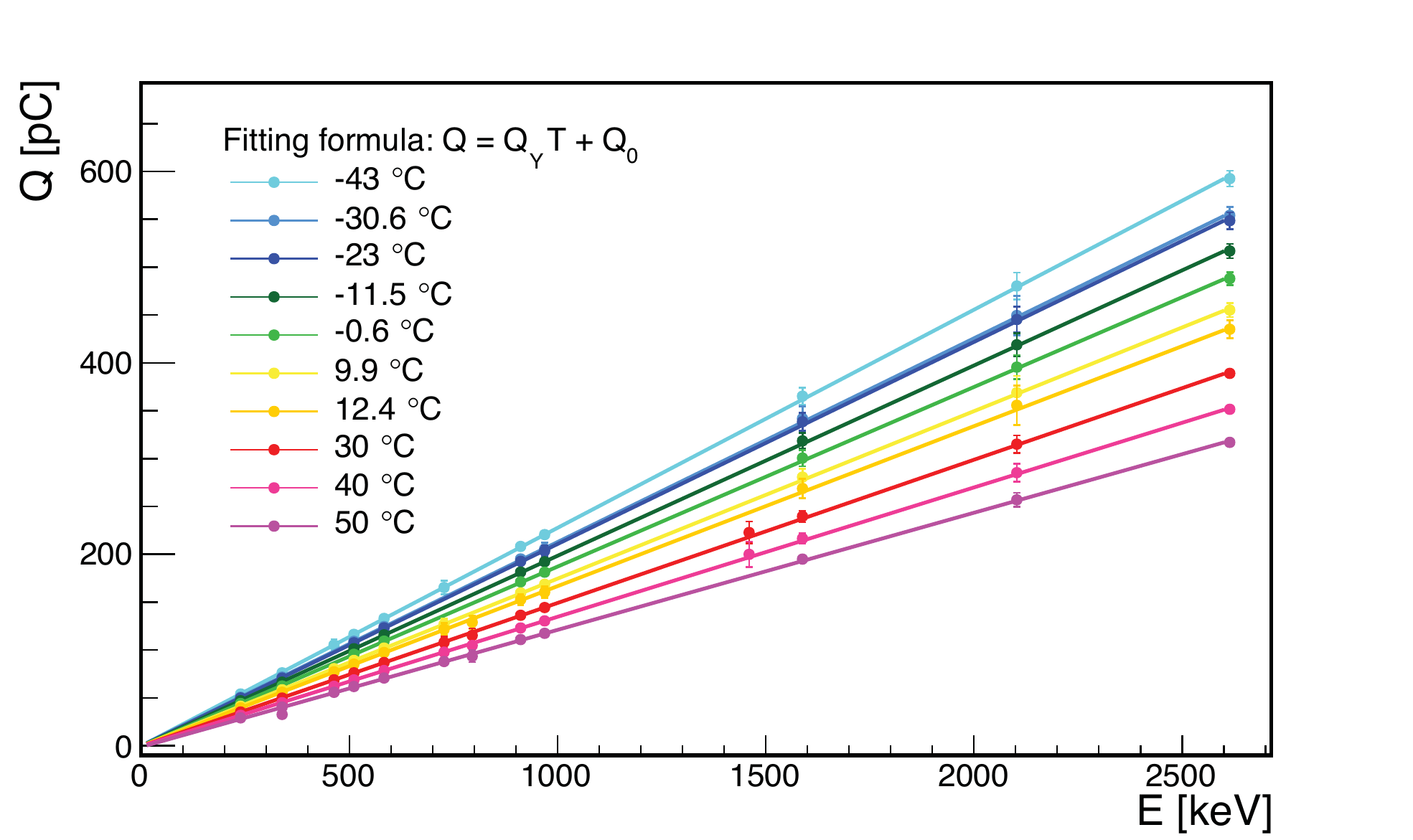}
    \caption{Energy calibration of the detector at different temperatures using $^{232}$Th.}
    \label{fig:Masdar_QvsE}
\end{figure}

\begin{figure}[t]
     \centering  
     \includegraphics[width=0.8\textwidth ]{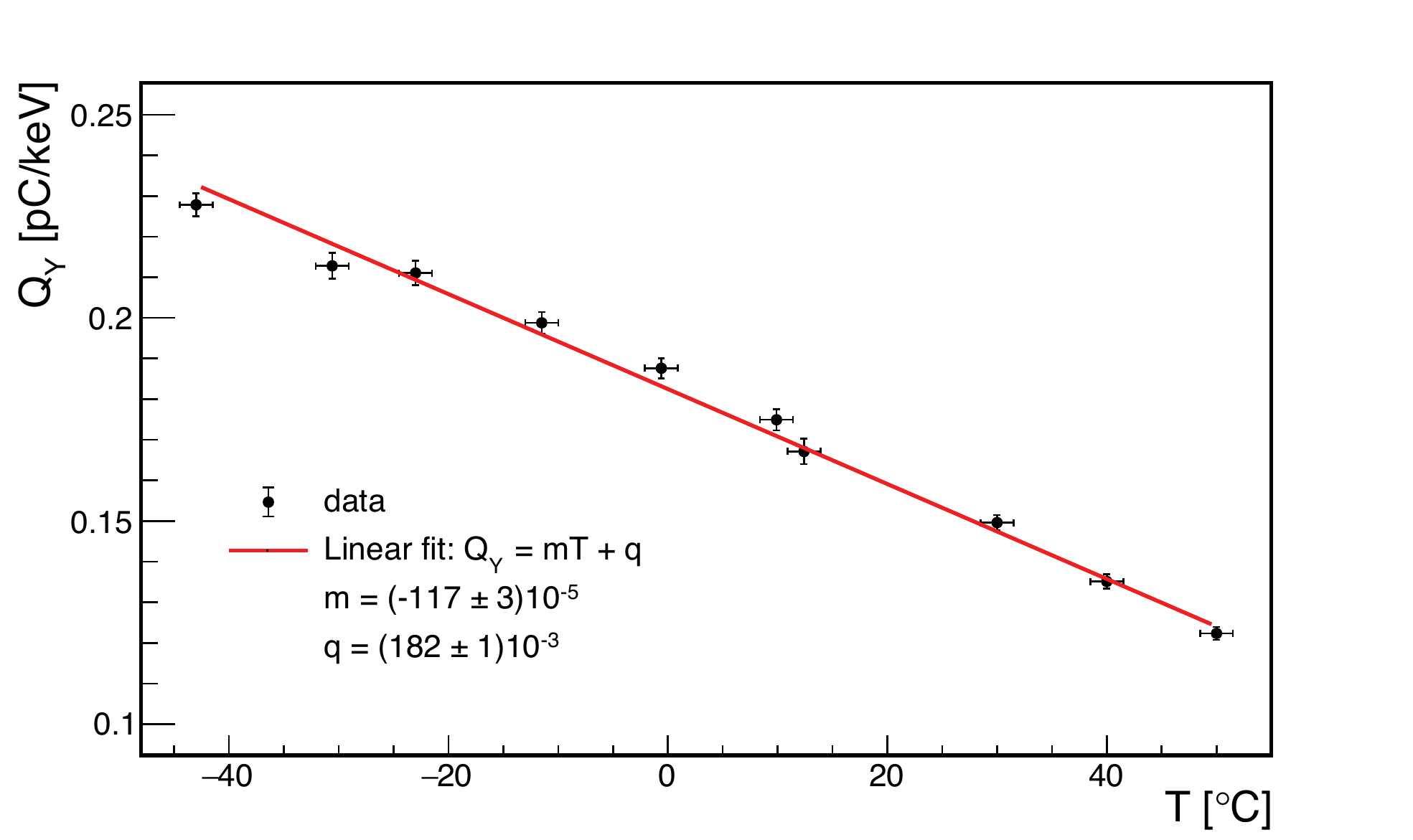}
     \caption{Charge yield in units of pC/keV as a function of temperature.}
     \label{fig:Masdar_slope}
 \end{figure}

By calibrating the detector at different temperatures using $^{232}$Th, we obtain the plot in Fig.~\ref{fig:Masdar_QvsE}, where each point represents the mean of the Gaussian fit of the corresponding energy peak in the charge spectrum at a given temperature, and the error bar is the standard deviation coming from the same fit.
Each data set in Fig.\ref{fig:Masdar_QvsE} is fitted with a linear function\footnote{Note that the offset $Q_0$ is not fixed to zero, but it turned out to be compatible with zero within uncertainties.} to get the charge yield in units of pC/keV as a function of temperature, shown in Fig.~\ref{fig:Masdar_slope}. Evaluating the function at \SI{23}{\celsius}, the temperature in the lab at NYUAD, gives a charge yield which is compatible within one standard deviation with the charge yield as measured at NYUAD in Fig.~\ref{fig:calibration}.
The two plots of Fig.~\ref{fig:Masdar_slope} and Fig.~\ref{fig:calibration} are obviously key to a proper energy reconstruction of events acquired in orbit. 

The dependence of the charge yield on the temperature can in principle be attributed to either the PMT or the crystal. While we do not have sufficient data to support either case, we know from the manufacturer \cite{HamamatsuHandbook} that the temperature coefficient of a bialkali photocathode is $-0.4 \%$\,$^\circ$C. Using this value, we expect a decrease of charge yield of $\approx$37\% from $-43^\circ$C to $50^\circ$C, while we observe a decrease of 46\%, indicating that most, if not all, of the variation is attributable to the PMT. A dedicated measurement will be needed to precisely disentangle the two contributions. 

\section{Conclusions}
\label{sec:conclusions}
We have fully characterised a prototype scintillation detector for its deployment on a CubeSat mission to study Terrestrial Gamma-ray Flashes. \cb\ has been chosen as scintillator for its excellent characteristics of speed, resolution, density, and low intrinsic radioactivity. The crystal has been coupled to a lightweight, compact photomultiplier, Hamamatsu R11265U-200, whose factory voltage divider has been replaced to meet the low power constraints of the mission. The unit has been fully calibrated in terms of gain (through single photoelectron measurements), energy,  linearity and resolution. 

Finally, we have presented the measurement of the temperature coefficient of the detector charge yield. A good detector characterisation as a function of temperature is a key factor for the space mission. In fact, to understand the physics behind TGFs the detector must be able to perform excellent spectroscopy, and it is therefore important to correct the signal amplitudes for temperature variations. 

\acknowledgments

The authors would like to thank M. Marisaldi from University of Bergen for the insightful comments that improved the quality of the work and L. Burderi from University of Cagliari for the fruitful discussions.

\bibliographystyle{JHEP}
\begingroup
    \setlength{\bibsep}{10pt}
    \bibliography{biblio.bib}
\endgroup

\end{document}